# Observation of the Critical State to Multiple-Type Dirac Semimetal Phases in KMgBi

D. F. Liu[1*], L. Y. W[2*], C. C. Le[3], H. Y. Wang[2], X. Zhang[2], N. Kumar[3], C. Shekhar[3], N. B. M. Schröter[4], Y. W. Li[2], D. Pei[4], L. X. Xu[2], P. Dudin[5], T. K. Kim[5], C. Cacho[5], J. Fujii[6], I. Vobornik[6], M. X. W[2,9], L. X. Yang[7,8], Z. K. Liu[2,9], Y. F. Guo[2], J. P. Hu[10], C. Felser[3], S. S. P. Parkin[1], Y. L. Chen[2,4,7,9†]

[1]*Max Planck Institute of Microstructure Physics, Halle, 06120, Germany*
[2]*School of Physical Science and Technology, ShanghaiTech University, Shanghai 201210, China*
[3]*Max Planck Institute for Chemical Physics of Solids, Dresden, D-01187, Germany*
[4]*Clarendon Laboratory, Department of physics, University of Oxford, Oxford, OX1 3PU, U.K.*
[5]*Diamond Light Source, Didcot, OX110DE. U.K.*
[6]*CNR-IOM, TASC Laboratory in Area Science Park, 34139, Trieste, Italy*
[7]*State Key Laboratory of Low Dimensional Quantum Physics, Department of Physics, Tsinghua University, Beijing 100084, China*
[8]*Frontier Science Center for Quantum Information, Beijing 100084, China*
[9]*ShanghaiTech Laboratory for Topological Physics, Shanghai 200031, China*
[10]*Institute of Physics, Chinese Academy of Sciences, Beijing 100190, China*

\* These authors contributed equally to this work.

† Corresponding author: yulin.chen@physics.ox.ac.uk

**Dirac semimetals (DSMs) are classified into different phases based on the types of the Dirac fermions. Tuning the transition among different types of the Dirac fermions in one system remains challenging. Recently, KMgBi was predicted to be located at a critical state that various types of Dirac fermions can be induced owing to the existence of a flat band. Here, we carried out systematic studies on the electronic structure of KMgBi single crystal by combining angle-resolve photoemission spectroscopy (ARPES) and scanning tunneling microscopy/spectroscopy (STM/STS). The flat band was clearly observed near the Fermi level. We also revealed a small bandgap of ~ 20 meV between the flat band and the conduction band. These results demonstrate the critical state of KMgBi that transitions among various types of Dirac fermions can be tuned in one system.**

# I. Introduction

Topological materials have been intensively studied recently, as they exhibit many unusual physical properties such as the existence of topologically protected dissipationless current and new types of fermions, such as Dirac, Weyl and Majorana fermions, that have great potential for applications in the electronic devices and quantum computing technology [1-4]. In the last several years, the success in the discovery of various topological materials by the combination between the theory and experiments, ranging from the topological insulators (TIs) [5-7], to Dirac semimetals (DSMs) [8-14] and Weyl semimetals (WSMs) [15-29], has greatly enriched the topological quantum states (TQSs) of matter.

Among these TQSs, DSMs have attracted intensive attention as they host linear Dirac cone in their electronic structures that can lead to many exotic properties [8,30-33]. According to the Fermi surface (FS) topologies, DSMs can be classified into three types: (i) type-I DSMs host point-like FS topology [10,11]; (ii) type-II DSMs host electron and hole Fermi pockets where the Dirac cone are strongly titled [12-14]; (iii) type-III DSMs appear at the phase transition point between type-I and type-II DSMs where the DP is formed by the crossing of a strict flat band and a linear dispersion with a line-like FS topology [34,35]. However, these three types of DSMs exist in different systems, i.e. type-I DSMs in $Na_3Bi$ [10] and $Cd_3As_2$ [11], type-II DSMs in $PtSe_2$ and $PtTe_2$ [12-14], while type-III DSMs were predicted in $Zn_2In_2S_5$ [34] and $Ta_2Se_8I$ [35], that brings difficulty in studying the topological phase transition among different types of Dirac fermions.

Recently, KMgBi was theoretically predicted to be located at a critical state that all the three types of Dirac fermions can be induced owing to the existence of a flat band [36]. The velocity sign of the flat band can be easily tuned by applying strain or substituting K atom by Rb or Cs, leading to the realization of various types of Dirac fermions [36]. KMgBi was also predicted to host one-dimensional topological hinge Fermi arcs connecting the bulk three-dimensional (3D) DPs [37]. Here, we carried out systematic studies on the electronic structures of KMgBi single crystal by combing angle-resolved photoemission spectroscopy (ARPES) and scanning tunneling microscopy/spectroscopy (STM/STS). We revealed the complete electronic structure of KMgBi and directly observed the flat band near the Fermi level. Moreover, a small bandgap of ~ 20 meV between the flat band and the conduction band was also observed. These results demonstrate the critical state of KMgBi that the transitions among various types of Dirac fermions can be tuned in one system.

## II. Methods

KMgBi single crystals were grown by using the self-flux method. The K chunk (99.5%), Mg chunk (99.9%) and Bi grains (99.9%) were mixed in the stoichiometric ratio and put into an aluminum crucible, and then sealed into a quartz tube with a partial pressure of argon. The assembly was slowly heated to 673 K, stayed at this temperature for 10 hours and then further heated up to 1123 K with staying for 2 days, followed by slowly cooling down to 773 K at a temperature decreasing rate of 2 - 3 K/h. KMgBi single crystals with a typical size $3 \times 4 \times 0.3$ mm$^3$ were obtained. It should be noted that all manipulations, except the sealing and reaction, were carried out in an argon-filled glove box. The phase and quality of KMgBi were examined on a Bruker D8 Venture single crystal X-ray diffractometer (SXRD) with Mo $K_{\alpha 1}$ ($\lambda$ = 0.71073Å) at 298 K. The ARPES experiments were performed at beamline I05 of the Diamond Light Source (DLS) with a Scienta R4000 analyzer and beamline APE of the Elettra synchrotron with a Scienta DA30 analyzer. The sample temperature was kept as 8K and 15 K in APE and DLS, respectively. The pressure were better than $2\times10^{-10}$ Torr. The angle resolution was <0.2º and the overall energy resolution was <15 meV. The STM/STS measurement was carried out in an ultra-high vacuum chamber with the pressure better than $2\times10^{-10}$ Torr. The KMgBi sample was cleaved in-situ in the preparation chamber at room temperature, then transferred to the STM sample stage kept at 5.2K. A PtIr tip was used for the imaging and tunneling, which has been calibrated on the sliver islands grown on p-type Si(111) 7×7 by means of molecule beam epitaxy (MBE). All the dI/dV curves were obtained by lock-in technique with a 5mV modulation at 971.2Hz. The samples were cleaved along the (001) plane for both ARPES and STM experiments. The band calculations were performed based on the projector augmented wave method encoded in the Vienna ab initio simulation package (VASP) code [38-40]. The Perdew-Burke-Ernzerhof (PBE) exchange-correlation functional and the projector-augmented-wave (PAW) approach are used for the exchange correlation potential. The cutoff energy is set to be 500 eV for expanding the wave functions into a plane-wave basis. In the calculation, the BZ is sampled in the momentum space within a Monkhorst-Pack scheme [41]. On the basis of the equilibrium structure, the momentum mesh used is 10 × 10 × 6. The calculations with a hybrid functional (using HSE03) [42] were also performed to study the ground state of the KMgBi. The amount of exact Hartree-Fock exchange is set to 0.17.

### III. Experimental results

The crystal structure of KMgBi is shown in Fig. 1a. It crystallizes in a tetragonal structure with the space group of *P4/nmm* and is comprised of ...K-Bi-$Mg_2$-Bi-K... layers (Fig.

1a). In each layer, the Mg atoms form a 2D square lattice with K and Bi atoms located alternatively above and below the Mg square (Fig. 1a). The distance of K and Bi atoms with respect to Mg plane are 2.927 Å and 1.724 Å, respectively (Fig. 1a). The adjacent K-Bi-Mg$_2$-Bi-K layers are weakly coupled by the Van der Waals forces, resulting in that the crystal naturally cleaves between K-K bonds that is ideal for the ARPES and STM/STS measurements.

The band structure of KMgBi is dominated by Bi-6$p$ orbitals near the Fermi level and the dispersions were calculated using two different methods as shown in Fig. 1c and 1d, respectively. Both calculations reveal a flat band (labeled as α) near the Fermi level along ΓZ direction. The flat band is dominated by Bi-6$p_{x,y}$ orbital and originates from the weak interaction between the two adjacent K-Bi-Mg$_2$-Bi-K layers along z direction. GGA calculations show the flat band is linearly intersected by the conduction band (labeled as β) and the type-I Dirac fermion is formed (see the inset in Fig. 1c), while HSE calculations reveal a small bandgap of 16.3 meV between the flat band and the conduction band without band inversion (see the inset in Fig. 1d). By applying strain or substituting K atom by Rb/Cs atom, such small bandgap can easily vanish, leading to the formation of the Dirac fermion [36]. In the meantime, the velocity sign of the flat band can also be tuned, driving the system into type-II DSM [36]. At the phase transition point between type-I and type-II DSM phases, type-III DSM phase can be realized where the velocity of the flat band strictly equals to zero.

To investigate the electronic structure of KMgBi, we first carried out ARPES measurements on the high-quality of KMgBi single crystals. Fig. 1e shows the sharp characteristic core levels of K, Mg and Bi (Fig. 1e). The broad FS mapping (Fig. 1f) covering multiple Brillouin zones (BZs) illustrates the overall FS topology of KMgBi, consisting of point-like FS at the center of BZ.

Fig. 2a illustrates the 3D volume plot of the electronic structure showing the point-like FS at $\bar{\Gamma}$ is formed by the top of a hole band. The large-energy scale of the band dispersion along different high symmetry directions across the whole BZ are shown in Fig. 2b. As the overall electronic structure derived from GGA (Fig. 1c) and HSE (Fig. 1d) calculations are consistent with each other except for the relative positions of the α and β bands, to understand the experimental band structure, we performed GGA calculations and the calculated band dispersion (red curves in Fig. 2b) are overlaid on the experimental results showing excellent agreements. Fig. 2c plots the experimental constant energy contours at different energies showing that the point-like FS (Fig. 2c (i)) gradually evolves into a four-petal-like feature (Fig. 2c (ii-iv)) with decreasing the energy to -0.6 eV (Fig. 2c (iv)). Such evolution can also be well reproduced by the calculations as shown in Fig. 2d.

According to the GGA and HSE calculations, the flat band (α band) is located along $k_z$ direction as illustrated in Fig. 3a (i) and 3a (ii) respectively. To search for the flat band, we performed detailed photon energy dependent measurement (20 to 90 eV) along $k_z$ direction as shown in Fig. 3b. The spectra intensity map at Fermi level in $k_y$-$k_z$ plane (Fig. 3b) shows a vertical line-like FS along ΓZ direction. Fig. 3c (i) and (ii) are the extracted band dispersions along ΓZ direction measured by linear horizontal (LH) and linear vertical (LV) polarizations of the photons, respectively. Under LH polarization (Fig. 3c (i)), the γ band located within energy range of -0.5 to -1 eV was clearly resolved (see the red dashed curves in Fig. 3c (i)), which is consistent with the calculations (Fig. 1c, d); while it was suppressed under LV polarization (Fig. 3c (ii)) due to the matrix element effect. Near the Fermi level, a nondispersive flat band was observed under both polarizations (Fig. 3c) as illustrated in Fig. 3a.

The flat band can also be visualized from the dispersions at different $k_z$ from Γ to Z points as shown in Figs. 3d and 3e. The VB1 band's top lies near the Fermi level and the positions under both polarizations are unchanged upon varying $k_z$ (Fig. 3d and 3e), as expected for the flat band. At energy of -0.6 eV, the VB2 band was only observed under LH polarization (Fig. 3d (i)) due to the matrix element effect. Its top position gradually moves down (Fig. 3d (i - vi)) forming the γ band (Fig. 3c (i)) upon varying $k_z$ from Γ to Z points. Fig. 3f shows the EDCs extracted at different $k_z$ along ΓZ direction. The peak position does not change upon varing $k_z$, again suggesting the dispersionless nature of the flat band.

GGA calculations show KMgBi is a DSM (Fig. 1c), while HSE calculations show it is a semiconductor (Fig. 1d). In a DSM, the density of state (DOS) is finite and becomes minimal at the DP; while it has a zero DOS at Fermi level due to the formation of a full bandgap in a semiconductor. To determine the ground state of KMgBi, we performed high-resolution STM/STS measurement as shown in Fig. 4. The large-scale topography map (Fig. 4a) exhibits inhomogeneous local DOS on the surface of the cleaved KMgBi sample. The STS spectra on the high DOS region (marked by the purple line in Fig. 4a) were systemically studied as shown in Fig. 4b. A gap structure is clearly observed at the Fermi level, characterized by the low intensity of the dI/dV spectra. We also studied the STS spectra on the low DOS region (Fig. 4c) and a similar gap structure was also observed (Fig. 4d). The large variation of the dI/dV spectra (Fig. 4b and 4d) maybe caused by the impurities and defects; for example the intrinsic Bi impurities and K vacancies, as well as the possible oxidization since KMgBi is very sensitive. To quantitatively determined the gap size, we selected several points on the sample surface (red and green circles in Figs. 4a and 4c) and the corresponding STS spectra were shown in Fig. 4e. Although the dI/dV spectra show large variation, all the dI/dV spectra show a bulk bandgap of ~ 20 meV, demonstrating the semiconducting state of KMgBi. Our results are consistent with the HSE calculations and the transport measurement [43].

## IV. Discussion

The inaccuracy of the GGA calculations in describing the ground state of KMgBi is caused by the inadequacy in describing the energy gap size between the valence and conduction bands due to the existence of a derivative discontinuity of the energy with respect to the number of electrons [44]. The HSE calculations with inclusion a certain amount of Hartree-Fock exchange can well capture the band gap of KMgBi. According to the HSE calculations, such small bandgap can be easily manipulated to drive KMgBi into DSM phases with different types of Dirac fermions [36], i. e. KMgBi can be turned into type-I DSM by applying out-of-plane tensile strain, while type-II DSM can be induced by applying in-plane compressive strain or by doping Rb or Cs; type-III DSM can also be realized at the topological phase transition point between type-I and type-II DSM phases. We note a resistivity plateau was observed in the transport measurement that was attributed to the existence of a nontrivial topological surface state [43]. However, all the observed electronic structures in our experiments are bulk states without any indications of surface state indicating that the resistivity plateau is more likely caused by the impurity.

## V. Conclusions

In summary, we have performed systematic studies on the electronic structure of KMgBi single crystals by combining ARPES and STM/STS. The flat band and a small bandgap of ~ 20 meV between the flat band and the conduction band were clearly observed. These results demonstrate the critical state of KMgBi that the transitions among various types of Dirac fermions can be tuned in one system.

DATA   AVAILABILITY

The data that support the findings of this study are available within the article.

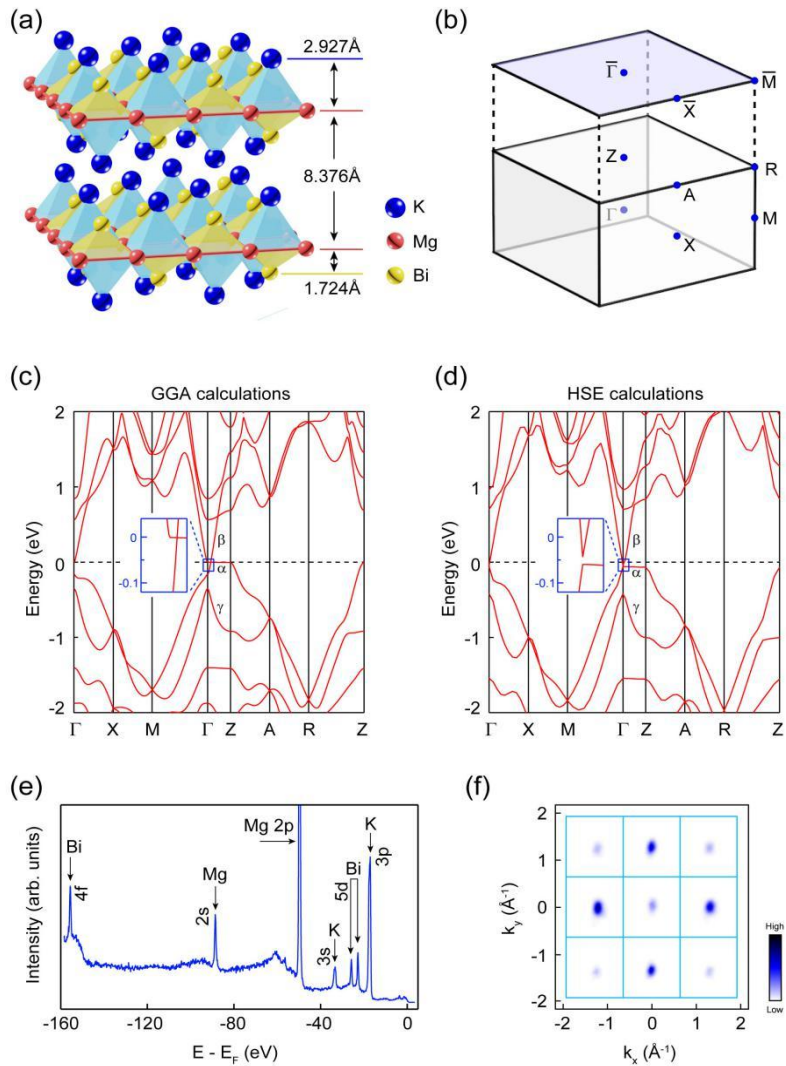

**Figure 1. The calculated electronic structure of KMgBi.** (a) Schematic illustration of the crystal structure of KMgBi. (b) The 3D BZ of KMgBi and its projected surface BZ to (001) plane. (c - d) Calculated bulk electronic structure of KMgBi with SOC based on the GGA (c) and HSE (d) calculations. Both calculations show a flat band (labeled as α) lying near Fermi level along ΓZ direction. (e) The core-level photoemission spectrum of KMgBi shows sharp characteristic peaks of K, Mg and Bi atoms. (f) Broad FS map covering multiple BZ shows the point-like FS at the center of BZ. The FS was taken by 150eV photon.

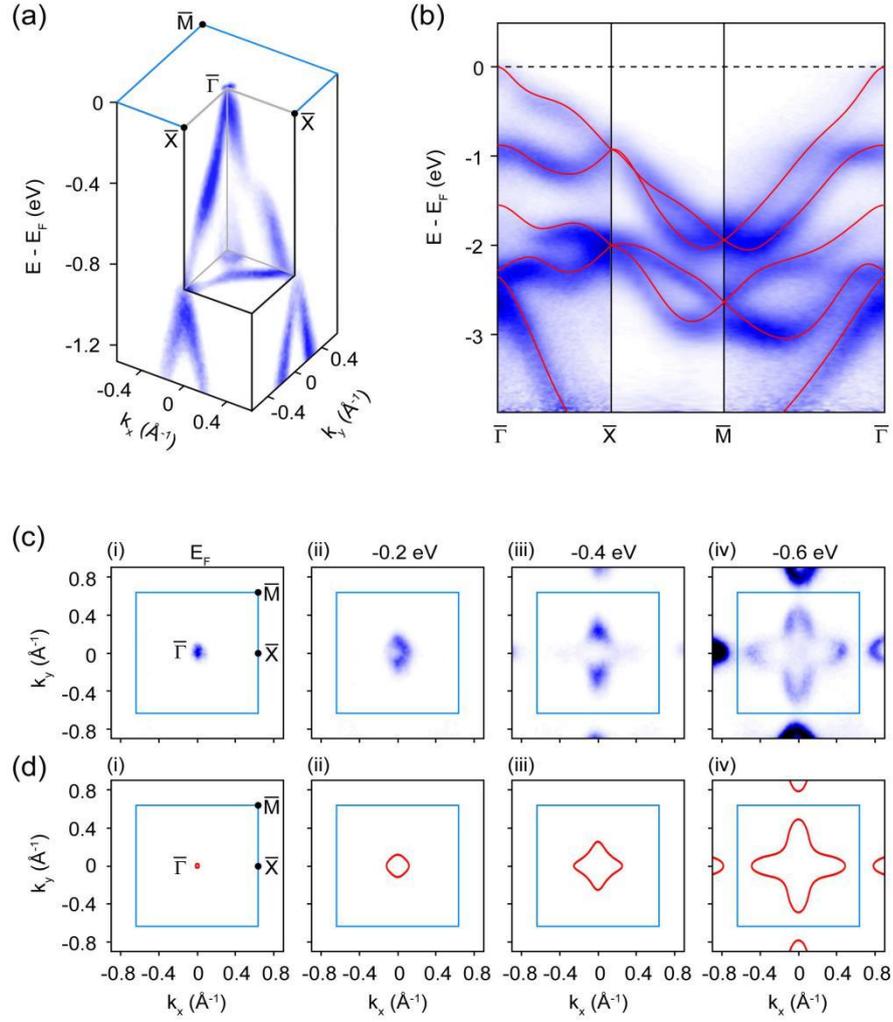

**Figure 2. The overall electronic structure of KMgBi.** (a) The 3D plots of the electronic structure. (b) Comparison of the experimental and calculated (red curves) band dispersions along high-symmetry directions across the whole BZ, showing excellent agreement. (c - d) The comparison of the experimental (c) and calculated (d) constant energy contours at different energies, showing excellent agreement. The calculated band structure was based on the GGA calculations and the bandwidth was renormalized by a factor of 0.9. The data was taken by 60eV photon.

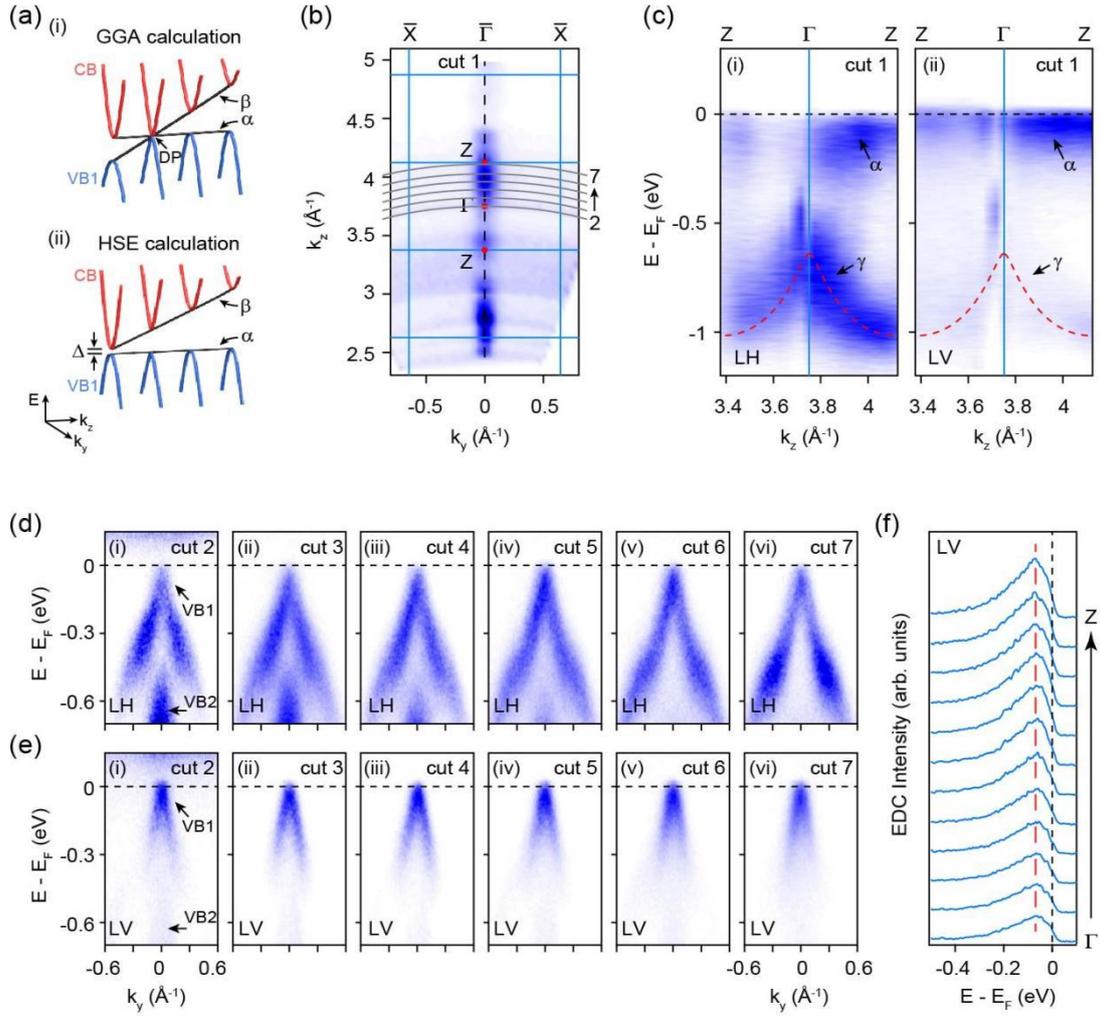

**Figure 3. Observation of the flat band.** (a) Schematic illustration for the formation of the flat band (labeled as α) based on the GGA (i) and HSE (ii) calculations. (b) The spectra intensity map at Fermi level in $k_y$-$k_z$ plane was taken by 20 - 90 eV photons under LH polarization. (c) The extracted band dispersion along ΓZ direction under LH (i) and LV (ii) polarizations. (d) The band dispersion at different $k_z$ from Γ to Z points measured under LH polarization. The corresponding $k_z$ is shown by the gray line in (b). (e) The same to (d) but measured under LV polarization. (f) The extracted EDCs at different $k_z$ (3.77 - 4.1 Å$^{-1}$) along ΓZ direction. The red dashed line is the guideline for the peak position.

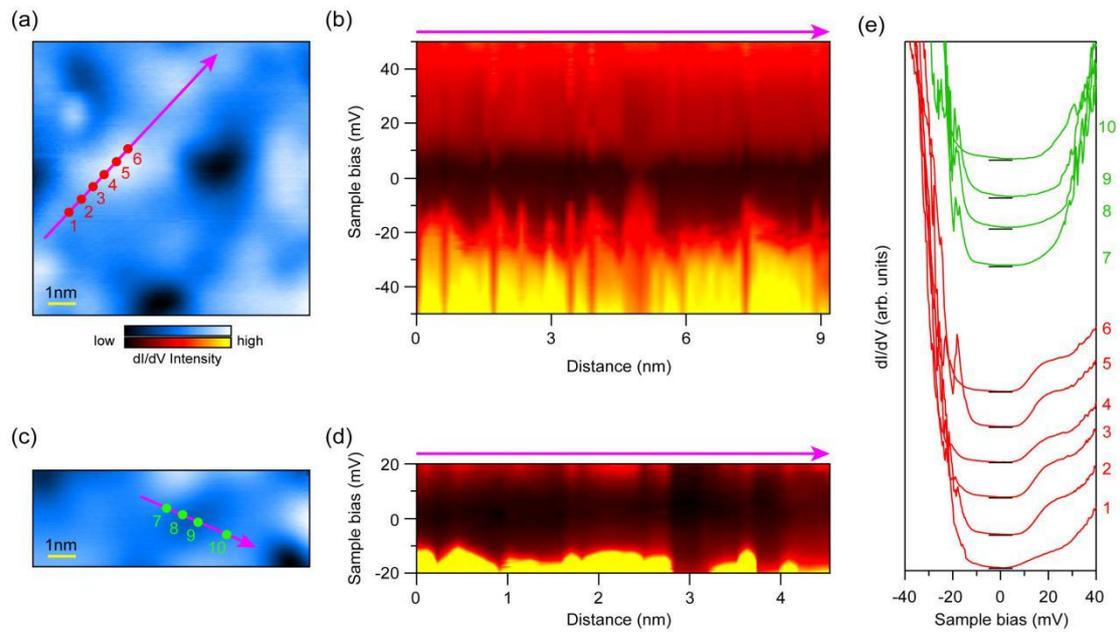

**Figure 4. Observation of a small semiconducting gap in KMgBi.** (a) Large-scale scan on the surface of the cleaved KMgBi single crystal showing inhomogeneous local DOS. (b) STS spectra on the high DOS region illustrated by the purple line in (a). (c) The STM image of the low DOS region. (d) STS spectra on the low DOS region illustrated by the purple line in (c). (e) STS spectra taken on several surface locations. The red curves were taken on the high DOS region and the corresponding positions are illustrated by the red circles in (a). The green curves were taken on the low DOS region and the corresponding positions are illustrated by the green circles in (c). The STS spectra are offset for charity. The black bars are the zero dI/dV intensity for each STS spectra. They all show a full gap of ~ 20 meV at the Fermi level, demonstrating the semiconducting state of KMgBi.